\documentstyle[11pt]{article}
\newcommand{\Lslash}[1]{ \parbox[b]{1em}{$#1$} \hspace{-0.8em}
                         \parbox[b]{0.8em}{ \raisebox{0.2ex}{$/$} }    }

\newcommand{\beq}{\begin{equation}}
\newcommand{\eeq}{\end{equation}}
\newcommand{\beqa}{\begin{eqnarray}}
\newcommand{\eeqa}{\end{eqnarray}}
\newcommand{\skipfields}{\!\!\!\!\! & \!\!\!\!\! &}
\begin{document}
\begin{center}
{\bf   The Chiral Dirac Determinant\\
\vspace{0.5cm}
According to the Overlap Formalism}\\
\vspace{1cm}
Per Ernstr{\" o}m
and
Ansar Fayyazuddin
\\
NORDITA, Blegdamsvej 17, DK-2100 Copenhagen {\O}, Denmark
\end{center}

\vspace{.25cm}
\begin{abstract}
The chiral Dirac determinant is calculated using the overlap formalism
of Narayanan and Neuberger.  We compare the real and imaginary parts
of the determinant with the continuum result for perturbative gauge
field backgrounds and show that they are identical.  Thus we
find that the overlap formalism passes a crucial test.
\end{abstract}
\vspace{0.25cm}

Nordita 95/80

November, 1995

\vspace{0.50cm}

Lattice regularization of field theories is the only known non-perturbative
regularization available to us.  Chiral gauge theories have eluded a
non-perturbative regularization for reasons summarized by the
Nielsen-Ninomiya \cite{no-go,shamir}
theorem which states that there exists no discretization
of the chiral Dirac operator which simultaneously
preserves a number of desirable
physical properties.  This is an unfortunate state of affairs since, at
least at low-energies, the Weinberg-Salam model describes the physics
of the world we live in, and this model involves chiral couplings of
fermions to gauge fields.  Recently attempts have been made to
evade the theorem of Nielsen and Ninomiya in various ways
(see \cite{shamir,ovrev} for a recent review of progress in this direction.)
We will be concerned with the approach of Narayanan and Neuberger\cite{nn} who,
inspired by an idea of Kaplan's\cite{kaplan},
have proposed a new way of calculating
chiral quantities on the lattice.  They evade the Nielsen-Ninomiya theorem
by studying an auxiliary problem in one dimension higher.  Quantities
in this auxiliary problem can then be related to the lower dimensional
theory by taking certain limits.  Thus the conclusions of the no-go theorem
are avoided
by formulating a problem which ostensibly has nothing to do with
the original problem and in fact is formulated in odd dimensions where
chirality is not an applicable concept.

In this letter we calculate the
determinant of the chiral Dirac operator in $4$
dimensions using the recipe of Narayanan and
Neuberger which has been dubbed the ``overlap formalism''.
First we evaluate the modulus of the determinant and show
that it reproduces correctly the continuum result.  We then evaluate
the phase of the determinant and
compare our result with the continuum result of Alvarez-Gaume, Della Pietra
and Della Pietra\cite{add}
(similar expressions have also been derived by \cite{ns,ball}.
We find that the results of the two approaches are
identical.
While calculating
the imaginary part of the effective action we  always work with
perturbative gauge fields since this is
the assumption under which the continuum results have been obtained.
Our results confirm that at least the chiral determinant can be defined
on the lattice using the overlap formalism.

The overlap formalism has been applied by a number of authors to various
problems involving chiral and non-chiral fermions.  Randjbar-Daemi and
Strathdee have calculated chiral anomalies in 2 and 4 dimensions, the
gravitational anomaly in 2 dimensions and the vacuum polarization
in 4 dimensions \cite{rspl,rsd,rs,rspol}.  They have also calculated
the two point functions for chiral fermions \cite{rspl} and verified
anomaly cancellation in the standard model using this formalism.
Randjbar-Daemi and Fosco  calculated the determinant of the chiral
Dirac operator in a constant background gauge field with non-trivial
holonomy on the two dimensional torus, verifying that the continuum
result is reproduced including the holomorphic anomaly \cite{rf}.
Narayanan and Neuberger have applied their formalism to the twisted chiral
Dirac operator and confirmed numerically the continuum result \cite{nnt}.
Narayanan, Neuberger and Vranas \cite{nnv} applied the overlap formalism to
the Schwinger model and obtained results consistent with the
continuum exact solution.  This last piece of work is particularly
interesting in that the gauge fields involved in that calculation
are topologically non-trivial and therefore involve zero modes of the
Dirac operator.
While work on the present project was in progress we received \cite{ks}
where the phase of the chiral determinant is calculated for
domain wall fermions.  The authors suggest that their results also
apply to the overlap formalism.

The overlap formalism expresses chiral quantities in terms of certain
objects in an auxiliary problem in one dimension higher.  Specifically,
one considers two five dimensional Hamiltonians (we use the simpler
notation of Randjbar-Daemi and Strathdee developed in \cite{rspl} and
\cite{rs})
\beq
H_{\pm} = \int d^{4}x \psi^{\dagger}\left( x\right)\gamma_{5}
\left(\Lslash D \pm\Lambda\right)\psi\left( x\right)
\eeq
where $\Lambda > 0$ is a mass for the five dimensional fermions.
Notice that the two Hamiltonians differ only by the sign of the mass term.
The Dirac vacua for the two Hamiltonians are denoted by $\mid A\pm>$ which
are Slater determinants of the non-positive eigenvalue states of the
first quantized Dirac Hamiltonians.
The overlap formalism states that the chiral Dirac determinant
is given by the following expression:

\beq
\mbox{det}[{{1}\over{2}}(1+\gamma_5 )(\Lslash D)]
= \lim_{\Lambda\rightarrow\infty} \frac{\langle+\mid A+ \rangle\langle A+\mid
A-
\rangle\langle A-\mid - \rangle}
{\langle+\mid - \rangle\mid \langle+\mid A+ \rangle\langle -\mid A-
\rangle\mid}
\label{eq:chiraldet}
\eeq
where $\mid\pm>$ are the Dirac vacua of the problem with vanishing gauge
fields.

Before proceeding to the calculation we comment on the regularization
procedure.
We assume that the five dimensional overlap problem can be regularized on
the lattice.
We will never explicitly define a lattice regularization but will assume
its existence and other formal properties of the lattice regularized
operator which must be identical with the continuum problem.
The Dirac operator as we will use it will always be a finite dimensional
matrix and we will take
\footnote{
For our calculation of the modulus of the determinant it is sufficient
to take $\Lambda \gg ||A||$
}
$\Lambda \gg 2\pi/a + ||A||$ where $||A||$ is the supremum of the expectation
value of the gauge field and
$a$ is the lattice spacing, ensuring that the eigenvalues of the
Dirac operator are {\em small} compared to the mass $\Lambda$.
There are two large scales in the problem: $\Lambda$ and the inverse
of the lattice spacing $1/a$.
Both are large but the relevant limit to reproduce the continuum
result is the one in which
$\Lambda \gg 1/a$.
However, one could imagine various limits controlled by the dimensionless
parameter $a\Lambda$.
We will have nothing to say about this  but we hope to return to this
issue in the future (see, however, the discussion in \cite{rs}\footnote{
After submitting this work we received a paper by Randjbar-Daemi and
Strathdee where this question is taken up (eprint archive: hep-th/9512112)}).
It would be very interesting to characterize any gauge non-invariant terms,
which we neglect in the present work, ``suppressed'' by $a\Lambda$
and others suppressed by $a$ and $1/\Lambda$ .
This may shed light on the continuum limit of the lattice regularized
overlap formalism.

The four dimensional Dirac operator anti-commutes with $\gamma_5$, this
allows one to pair the non-zero eigenmodes of the operator as follows:
\beqa
{\Lslash D}\left( A\right)\phi_{j}\left( A\right) & = &
i\epsilon_{j}\left( A\right)\phi_{j}\left( A\right), \\
{\Lslash D}\left( A\right)\gamma_{5}\phi_{j}\left( A\right)&  = &
-i\epsilon_{j}\left( A\right)\gamma_{5}\phi_{j}\left( A\right).
\eeqa
We adopt the notation that the $\phi_{j}$ are {\em positive} eigenvalue
modes of $-i{\Lslash D}$ and $\epsilon_{j}$ denotes a positive eigenvalue.
We assume first that there are no zero modes.
Then the eigenstates
$\left\{\phi_{j}, \gamma_{5}\phi_{j}\right\}$
of the Dirac operator form a complete basis.
%
%
We will always assume that $\phi_{j}\left( 0\right)$ and
$\phi_{j}\left( A\right)$ are smoothly related to each other by an
interpolating gauge field $A_{t}$ between the gauge configurations
$0$ and $A$ such that the Dirac operator does not develop a zero mode
anywhere along the interpolation.
The first quantized hamiltonians:
\beq
h_{\pm} = \gamma_{5}\left(\Lslash D \pm\Lambda\right)
\eeq
commute with ${\Lslash D}^{2}$ and allow one to diagonalize
$h_{\pm}$ and ${\Lslash D}^2$ simultaneously.
We can express the eigenstates of $h_{\pm}$ as linear combinations
of a positive eigenstate of $\Lslash D$ and its negative eigenvalue
pair.  This has the virtue that the eigenstates
of $h_{\pm}$ will be linear combinations of $\phi_{j}, \gamma_{5}\phi_{j}$
and the $\Lambda$ dependence will occur only in the coefficients
multiplying the eigenstates of ${\Lslash D}$.
Using the usual notation denoting positive (negative) eigenstates
of the hamiltonians $h_{\pm}$ as $u_{\pm}$($v_{\pm}$) we get the
following eigenstates:
\beqa
v_{\pm,j}\left( A\right) & = &\frac{1}{\sqrt{2}}\left(1-
\frac{i\epsilon_{j}\left( A\right)\pm\Lambda}
{\sqrt{\epsilon_{j}^{2}\left( A\right)+\Lambda^{2}}}\gamma_{5} \right)
\phi_{j}\left( A\right),\nonumber \\
u_{\pm,j}\left( A\right) & = &\frac{1}{\sqrt{2}}\left(1+
\frac{i\epsilon_{j}\left( A\right)\pm\Lambda}
{\sqrt{\epsilon_{j}^{2}\left( A\right)+\Lambda^{2}}}\gamma_{5}\right)
\phi_{j}\left( A\right).
\label{eq:wave}
\eeqa
These wave functions satisfy:
\beqa
h_{\pm}v_{\pm,j}& =
&-\sqrt{\epsilon_{j}^{2}+\Lambda^{2}}v_{\pm,j},\nonumber \\
h_{\pm}u_{\pm,j}& = &+\sqrt{\epsilon_{j}^{2}+\Lambda^{2}}u_{\pm,j}.
\eeqa
The Dirac vacua are then given by the Slater determinants of
the negative energy states.  We have normalized the above wave
functions in such a way that if we have an inner product such that
\beqa
& &\left(\phi_{j}\left( A\right),\phi_{k}\left( A\right)\right)
= (\gamma_{5}\phi_{j}\left( A\right),\gamma_{5}\phi_{k}\left( A\right))
= \delta_{jk}, \nonumber \\
& &(\phi_{j}\left( A\right),\gamma_{5}\phi_{k}\left( A\right)) = 0.
\label{eq:orth}
\eeqa
then the
$v_{\pm}, u_{\pm}$ are orthonormal with respect to the same inner
product.

Now we will calculate $\langle A+\mid A- \rangle$.  The states $\mid A\pm>$ are
Slater determinants of the $v_{\pm}$,  assuming for the moment that
$\Lslash D$ has no zero modes,
\beqa
\langle A+\mid A- \rangle& = &\mbox{det}M, \nonumber \\
M_{jk} &=& (v_{+,j},v_{-,k}).
\eeqa
We can now evaluate $M_{jk}$ using equations (~\ref{eq:wave}) and
(~\ref{eq:orth}).  We find
\beqa
M_{jk}& = &\delta_{jk}\frac{\epsilon_{j}\left( A\right)}{\epsilon_{j}
\left( A\right) + i\Lambda}, \nonumber \\
\mbox{det}M & =&\prod_{j}\frac{\epsilon_{j}\left( A\right)}
{\epsilon_{j}\left( A\right)+i\Lambda}.
\eeqa
Thus
\beqa
\lim_{\Lambda\rightarrow\infty}\frac{\langle A+\mid A- \rangle}{\langle+\mid -
\rangle} &=&
\lim_{\Lambda\rightarrow\infty}
\prod_{j}\frac{\epsilon_{j}\left(A\right)}{\epsilon_{j}\left(A\right)+i\Lambda}
\frac{\epsilon_{j}\left(0\right)+i\Lambda}{\epsilon_{j}\left(0\right)}
\nonumber \\
&=& \prod_{j}\frac{\epsilon_{j}\left(A\right)}{\epsilon_{j}\left(0\right)} \\
&=& \sqrt{\left|\frac{\mbox{det}\Lslash D\left( A\right)}
{\mbox{det}\Lslash\partial}\right|}.
\label{detmodulus}
\eeqa
The last equality follows from recalling that the $\epsilon_{j}$ are the
positive eigenvalues of $\Lslash D$.

Now we would like to show that if $\Lslash D$ has zero modes then
$\langle A+\mid A- \rangle=0$.  To demonstrate this we divide the
zero modes of
$\Lslash D$ into positive ($L$) and negative ($R$) ``chirality''
(i.e. with respect to $\gamma_{5}$) modes.
Then denoting by $\psi_{L(R),j}$ the zero modes of $\Lslash D$
we see that they are eigenstates
\footnote{Note that the index $j$ does not necessarily
run over the same number of values for the left and right handed wave
functions, this number can be different if the gauge field carries
a non-zero instanton number.}
of $h_{\pm}$:
\beqa
h_{\pm}\psi_{L,j}& = &\pm \Lambda\psi_{L,j}, \nonumber \\
h_{\pm}\psi_{R,j}& = &\mp \Lambda\psi_{R,j}.
\eeqa
Therefore, the right handed zero modes will be in the $\mid A+ \rangle$ vacuum
but will not appear in $\mid A- \rangle$.  The converse is true for the left
handed zero modes.  Using the orthogonality of the
left and right handed modes then proves that $\langle A+\mid A- \rangle=0$.
Of course, if the number of left and right handed zero modes is not
the same then $\langle A+\mid A- \rangle=0$ for an additional
reason than the one
just stated, namely, there is a mismatch in the number of states
in the two vacua.

So far we have evaluated $\langle A+\mid A- \rangle/\langle+\mid - \rangle$
for arbitrary gauge configurations.
Since the remaining part of the Dirac determinant (~\ref{eq:chiraldet})
is a phase while $\langle A+\mid A- \rangle/\langle+\mid - \rangle$ is a real
non-negative number we have evaluated the magnitude of the chiral Dirac
determinant and found that it is precisely as it should be.
We turn now to the phase of the Dirac determinant which is, in a sense,
at the heart of the matter since all the
information about chirality is stored in this phase.
The magnitude is merely the square root of the full Dirac determinant
with vector couplings.

We are interested in calculating the phase of the determinant in background
configurations for which there are continuum results available for
comparison.  The phase of the determinant was calculated by Alvarez-Gaume,
Della-Pietra and Della-Pietra for perturbative background gauge fields for
which there are no zero modes of the Dirac operator.  They found that
the phase can be written as:
\beq
\mbox{Im}\ln\mbox{det} = \pi\eta_{1}\left( 0\right) -
2\pi Q_{5}^{0}\left( {\hat A}_{t,1}\right).
\eeq
Where $Q_{5}^{0}\left[ {\hat A}_{t,1}\right]$ is the five dimensional
Chern-Simons
form and ${\hat A}_{t,u}$ is a two parameter extension of the four
dimensional
gauge field such that ${\hat A}_{t,u} = 0$
for $-\infty \leq t \leq -T$, ${\hat A}_{t,u}$
smoothly interpolates between $0$ and $A_{u}$ for $-T < t <T$
and finally ${\hat A}_{t,u}= A_{u}$ for $T < t <\infty$.
While $A_{u}$ is an interpolating field between $0$ (for $u=0$)
and $A$ (for $u=1$).  $A$ is the four dimensional gauge field
appearing in the Dirac operator
whose determinant we wish to calculate.  The object
$\eta_{u}\left( 0\right)$ is
the so-called eta-invariant associated with the five dimensional Dirac
operator
$H_{u}= i\gamma_{5}\partial_{t} + i{\Lslash D}\left( {\hat A_{t,u}}\right)$.
Crucial to the derivation of this result is the identity derived in \cite{add}:

\beq
\pi \frac{d\eta_{u}\left( 0\right)}{du} = 2\pi \frac{d}{du}Q_{5}^{0}
+\frac{1}{2}\mbox{tr}\gamma_{5}\left(\frac{d{\Lslash D}\left( A_{u}\right)}
{du}\right){\Lslash D}^{-1}\left( A_{u}\right)
\left( 1-{\Lslash D}^{2}\left( A_{u}\right)/M^{2}\right)^{-1/2}
\eeq
The $M$ appearing on the right hand side is a Pauli-Villars mass
regularizing the expression, and a limit where it is taken to infinity is
implicit.  If one takes this limit the last term on the
right hand side becomes:
\beqa
\lim_{M\rightarrow\infty}
\frac{1}{2}\mbox{tr}\gamma_{5}\left(\frac{d{\Lslash D}\left( A_{u}\right)}
{du}\right){\Lslash D}^{-1}\left( A_{u}\right)\skipfields
\left( 1-{\Lslash D}^{2}\left( A_{u}\right)/M^{2}\right)^{-1/2} \nonumber \\
\skipfields =\frac{1}{2}\mbox{tr}\gamma_{5}\left(\frac{d{\Lslash D}\left(
A_{u}\right)}
{du}\right){\Lslash D}^{-1}\left( A_{u}\right).
\eeqa
This expression, without the Pauli-Villars mass,  has appeared in the
physics literature previously
in a paper by Niemi and Semenoff \cite{ns}.
Alternatively, one could define a lattice
regularization of the operators appearing in the trace and then remove
the Pauli-Villars regulator.  In either case we need to take this limit
to be able to compare with our calculation which has no Pauli-Villars regulator
but a lattice regulator instead.
Finally, we can evaluate the trace over the basis $\left\{\phi_{j}
\left(A_{u}\right), \gamma_{5}\phi_{j}\left(A_{u}\right)\right\}$ to get
\beqa
\frac{d}{d u}\left(
\pi\eta_{u}\left( 0\right) - 2\pi Q_{5}^{0}\left( {\hat A}_{t,u}\right)
\right) & = &
\frac{1}{2}\mbox{tr}\gamma_{5}\left(\frac{d{\Lslash D}\left( A_{u}\right)}
{du}\right){\Lslash D}^{-1}\left( A_{u}\right)
\nonumber \\
 & = &
2\sum_{k}\left( \phi_{k}\left( A_{u}\right),
\gamma_{5}\frac{d}{du}\phi_{k}\left( A_{u}\right)\right).
\eeqa

The phase of the determinant in the overlap formalism is given
by
\beq
\mbox{Im}\ln \langle+\mid A+ \rangle\langle A-\mid - \rangle = \mbox{Im}\left(
\ln\mbox{det}M_{+}
+\ln\mbox{det}M_{-}\right).
\eeq
Where
\beqa
\langle+\mid A+ \rangle & = &\mbox{det}M_{+}, \nonumber \\
M_{+jk} & =& \left(v_{+j}\left( 0\right),v_{+k}\left( A\right)\right)
\nonumber \\
& = & \frac{1}{2}\left[\left( 1 + \frac{-i\epsilon_{j}\left( 0\right) +\Lambda}
{\sqrt{\epsilon_{j}^{2}\left( 0\right) +\Lambda^{2}}}
\frac{i\epsilon_{k}\left( A\right) +\Lambda}
{\sqrt{\epsilon_{k}^{2}\left( A\right) +\Lambda^{2}}}\right)
\left(\phi_{j}\left( 0\right),\phi_{k}\left( A\right)\right) \right.
\nonumber \\
& - &\left.\left( \frac{-i\epsilon_{j}\left( 0\right) +\Lambda}
{\sqrt{\epsilon_{j}^{2}\left( 0\right) +\Lambda^{2}}}
+ \frac{i\epsilon_{k}\left( A\right) +\Lambda}
{\sqrt{\epsilon_{k}^{2}\left( A\right) +\Lambda^{2}}}\right)
\left(\phi_{j}\left( 0\right),\gamma_{5}\phi_{k}\left( A\right)\right)
\right],
\eeqa
and
\beqa
\langle A-\mid - \rangle & = &\mbox{det}M_{-}, \nonumber \\
M_{-jk} & =& \left(v_{-j}\left( A\right),v_{-k}\left( 0\right)\right)
\nonumber \\
& = & \frac{1}{2}\left[\left( 1 + \frac{i\epsilon_{j}\left( A\right) +\Lambda}
{\sqrt{\epsilon_{j}^{2}\left( A\right) +\Lambda^{2}}}
\frac{-i\epsilon_{k}\left( 0\right) +\Lambda}
{\sqrt{\epsilon_{k}^{2}\left( 0\right) +\Lambda^{2}}}\right)
\left(\phi_{j}\left( A\right),\phi_{k}\left( 0\right)\right)\right.
\nonumber \\
& +&\left.\left( \frac{i\epsilon_{j}\left( A\right) +\Lambda}
{\sqrt{\epsilon_{j}^{2}\left( A\right) +\Lambda^{2}}}
+ \frac{i\epsilon_{k}\left( 0\right) +\Lambda}
{\sqrt{\epsilon_{k}^{2}\left( 0\right) +\Lambda^{2}}}\right)
\left(\phi_{j}\left( A\right),\gamma_{5}\phi_{k}\left( 0\right)\right)
\right].
\eeqa

We now take the limit $\Lambda\rightarrow\infty$ and evaluate the determinants
in that limit.
\beqa
M_{+jk} &\rightarrow& \left( \phi_{j}\left( 0\right),\left( 1-\gamma_{5}\right)
\phi_{k}\left( A\right)\right),  \nonumber \\
M_{-jk} &\rightarrow& \left( \phi_{j}\left( A\right),\left( 1+\gamma_{5}\right)
\phi_{k}\left( 0\right)\right).
\eeqa
To compare the overlap method with the continuum method
we consider a
field $A_{u}$ interpolating between the configurations
$A_{0}=0$ and $A_{1}= A$.  Using the completeness of $\left\{\phi_{j}(0),
\gamma_{5}\phi_{j}\left( 0\right)\right\}$
It is easy to check that $M_{\pm}$ are
unitary matrices in the limit $\Lambda \rightarrow \infty$.  Thus:
\beqa
\frac{d}{du}\ln\mbox{det}M_{\pm u}
& = & \mbox{tr}M_{\pm u}^{\dagger}\frac{d}{du}M_{\pm u} \nonumber \\
& = & \pm\sum_{k}\left(\phi_{k}\left( A_{u}\right),
\left(1\pm\gamma_{5}\right)\frac{d}{du}\phi_{k}\left( A_{u}\right)\right).
\eeqa
We arrive finally at the expression:
\beqa
\frac{d}{du}\ln\langle+\mid A+ \rangle\langle A-\mid - \rangle
& = &
 \frac{d}{du}\left(\ln\mbox{det}M_{+u}+\ln\mbox{det}M_{-u}\right)
\nonumber \\
& = &
2\sum_{k}\left(\phi_{k}\left( A_{u}\right),
\gamma_{5}\frac{d}{du}\phi_{k}\left( A_{u}\right)\right).
\eeqa
Comparing with equation (17) we see that the continuum result
for the phase of the chiral determinant
is reproduced by the overlap formalism.
Together with equation (\ref{detmodulus}) we have
\beq
\lim_{\Lambda\rightarrow\infty}
\frac{\langle+\mid A+ \rangle
      \langle A+\mid A- \rangle
      \langle A-\mid - \rangle}
     {\langle+\mid - \rangle \mid
      \langle+\mid A+ \rangle
      \langle -\mid A- \rangle \mid } =
\parbox[b]{22mm}{ \raisebox{-4mm}{${}^{
  \sqrt{ \left|\frac{\mbox{det}\Lslash D\left( A\right)}
                    {\mbox{det}\Lslash\partial}\right|}   }
$}}
e^{i\left(
\pi\eta_{1}\left( 0\right) - 2\pi Q_{5}^{0}\left( {\hat A}_{t,1}\right)
      \right)} .
\eeq

We conclude with a few comments.  The overlap recipe for the chiral Dirac
determinant has passed an important test by reproducing the continuum result.
What is most satisfying about this result is that while in the continuum
the imaginary part of the effective action naively vanishes but is
produced due to the regularization of the determinant and survives
the limit in which the regulator is removed, in the overlap formalism
one can reproduce this result independently of the specifics of
the regularization procedure
and any delicate limits.  This is a consequence of the fact that we
are always working with a parity non-invariant system
 which has a non-vanishing imaginary
part independently of the
lattice regulator.  In our derivation we have neglected terms of order
$\epsilon /\Lambda$ where $\epsilon$ is a typical eigenvalue of the Dirac
operator.  It would be interesting to keep these terms and to see how
various limiting procedures in which $a\rightarrow 0$ and
$\Lambda\rightarrow\infty$ can affect continuum limits.  In any problem
where more than one large scale is available one must eventually address
the question of their relative size.

Acknowledgements
We would like to thank B. Kileng, A. Krasnitz, L. K{\"a}rkk{\"a}inen,
and especially A. Kronfeld for discussions.
We are grateful to M. Schmaltz for pointing out some misprints in an
earlier version of this article.

\end{document}